# An Empirical Study of the Relationships between Code Readability and Software Complexity

Duaa Alawad[1], Manisha Panta[1], Minhaz Zibran[2], Md Rakibul Islam[3]
Department of Computer Science, The University of New Orleans
New Orleans, Louisiana, United States

**Abstract**

Code readability and software complexity are important software quality metrics that impact other software metrics such as maintainability, reusability, portability and reliability. This paper presents an empirical study of the relationships between code readability and program complexity. The results are derived from an analysis of 35 Java programs that cover 23 distinct code constructs. The analysis includes six readability metrics and two complexity metrics. Our study empirically confirms the existing wisdom that readability and complexity are negatively correlated. Applying a machine learning technique, we also identify and rank those code constructs that substantially affect code readability.

## 1. Introduction

Code Readability is defined as a human judgment as to how much source code is understandable and easy to read. It has been traditionally considered as an important software quality metric as it has a great influence on software maintenance. Typically, maintainability phase consumes 40% to 80% of the total life-cycle cost of a software [1]. Aggarwal et al. [2] claim that source code readability and documentation readability are crucial for maintainability of a software. Some researchers identify reading code as a key activity in software maintenance, and also recognize it as the most time-consuming activity among all the maintenance activities [3], [4], [5].

Software complexity is defined in IEEE glossary standards as: "the degree to which a system or component has a design or implementation that is difficult to understand and verify" [12]. The complexity of code can be affected by many factors, such as: lines of code, total number of operators and operands, coupling between objects, and number of control flows [13]. These factors are used in software metrics for measurement and approximate quantification of software complexity.

Researchers have established the importance of code readability and software complexity to the quality of software. Software complexity is considered as an "essential" property of the software since it reflects the complexity of the real-worlds problem a software deal with [6]. On the other hand, code readability is considered as an "accidental property", not an essential one, as it is not determined by the problem space, and can largely be controlled by the software engineers. While software complexity metrics measure the size of classes and methods, coupling, and interdependencies between modules, the code readability considers local and line-by-line factors such as: names of identifies, indentations, spaces, and length of lines of code.

Software quality is a critical topic in software engineering, and thus many researchers have performed studies in this area. Code readability and software complexity have a substantial impact on software quality. For better quality, low complexity and high readability are desired. Complexity may impact code readability, while low readability also may result in higher perceived complexity. Thus, readability and complexity are related.

**Research Questions:** A proper understanding of the relationship between these two attributes (i.e., readability and complexity) is necessary. In this paper, we present an empirical study of the relationships between code readability and software complexity. Using variety of readability and complexity metrics at a fine level of granularity, we examine the relationships between these two attributes. In particular, we aim to answer following two research questions:

***RQ1:*** *What type of relationship exists between Code Readability and Software Complexity?*

***RQ2:*** *What are the code constructs that affect Code Readability?*

The answers to the research questions are derived based on analyses of 35 Java programs covering 23 distinct programming constructs (e.g., loop, nested loop, conditionals). Our study derives empirical evidence to confirm the existing wisdom that code complexity and readability are inversely correlated. Applying a machine learning technique, we also identify code constructs that substantially affect code readability.

**Outline:** The remainder of this paper is organized as follows: In Section 2, we discuss research relevant to this study, especially related work on readability metrics and complexity metrics. In Section 3, we describe the setup of our study. Here we introduce the metrics, tools, and datasets used in our work. Section 4 includes the analyses, findings, our derived answers to the two research questions. In Section 5, we discuss the threats to the validity of this study. Finally, Section 6 concludes the paper with a summary and future research directions.

## 2. Related Work

In this section, we discuss research relevant to this study of ours. We mainly confine our discussion to the work that involved readability and complexity metrics the studies of their relationships.

### 2.1 Software Metrics

**Readability Metrics:** Researchers are working on determining code readability metrics and which specific code constructs influence readability of the software [7], [8], [9], [14], [15]. Buse & Weimer [9], [10] have proposed a code readability metric and developed a readability tool that automatically measures proposed readability metric. They selected Java code snippets and made them available to the selected human annotators for the judgement of readability of those code snippets. The results obtained from the experts were compared with results from the propose readability tool. The overall accuracy of the tool was found to be 80%. The study also showed that the readability is strongly correlated with some software quality attributes such as code changes, defect log messages and automated defect reports.

However, Daryl Posnett et al. [16] argued that code readability is a subjective property and it is not persuadable to generate readability score using automated readability tool. Further, they included that readability very much depends on the information contained in the source code and thus the readability score can be calculated based on size and code entropy. Similarly, Ankit et al. [15] performed a review of metrics for software quality. The authors reviewed various readability metrics in the literature such as Flesh-Kincaid metric, Gunning-Fog metric, SMOG index, Automated Readability Index and Coleman-Liau Index and concludes that the choice of readability metrics depend on different employments. They mention various elements of code that improves and degrades readability, for example appropriate comments and poorly defined variables respectively.

**Complexity Metrics:** Many software complexity metrics have been proposed in the past. Almost all the proposed complexity metrics measure complexity based on three attributes: software size, data flow, and control flow. Halstead Complexity Model [18], McCabe [19], Line of Code (LOC) [20] and Chidamber & Kemerer [21] metrics suites are all examples of proposed complexity metrics.

### 2.2 Analyses of Relationship

There have been few investigations into the relationship between software readability and complexity [7], [14], [27].

In the study of Puneet et al. [7], the investigation was based on Component Based Software Engineering (CBSE). The study uses a complexity metric that measures interface complexity for software components and shows how it is strongly correlated to readability for software components. The results indicate a negative correlation between readability and complexity.

Buse and Weimer [14] proposed an approach for constructing a readability tool. They investigated correlation between readability score from their proposed readability tool and cyclomatic complexity by using Pearson product moment correlation. They found that readability is weakly correlated with complexity in the absolute sense, and it is effectively uncorrelated to complexity in relative sense.

## 3. Study Setup

In this section, we discuss readability and complexity metrics used in this work, as well as the tools and methodology adopted in this research.

### 3.1 Readability Metrics

Different metrics are developed to estimate the readability of code. The readability metrics used in this work are described below.

*A. The Automated Readability Index (ARI):*
ARI [28] is based on the ratio of sentence difficulty to word difficulty. Sentence difficulty is determined as words per sentence and word difficulty is that by calculating letters per words. The equation for calculating ARI is:

$$ARI = 4.71(characters) + 0.5 (words) - 21.43$$

The numeric value of the ARI metric it approximates the grade level needed to comprehend the text. For example, ARI = 3 means, students in 3rd grade (ages 8-9 yrs. old) should be able to comprehend the text [28].

*B. The Simple Measure of Gobbledygook (SMOG):*
SMOG is suggested by G Harry McLaughlin [22] in 1969. This metric evaluated the time (in years) required by any

person to read the text. The equation for calculating SMOG is:

*SMOG = 3 + Square Root of Polysyllable Count*

The SMOG metric value signifies a U.S. school grade level indicating that an average student in that grade level can read the text [28]. For example, SMOG = 7.4 indicates that the text is understood by an average student in 7th grade.

C. *Flesch-Kincaid Readability Index (FKI):*
Flesch-Kincaid Readability Index [24] value is computed using the following formula:

$$206.835 - 1.015 \left(\frac{total\ words}{total\ sentences}\right) - 84.6 \left(\frac{total\ syllables}{total\ words}\right)$$

High value of Flesch-Kincaid Index (FKI) indicates high code readability whereas low value implies code is hard to read. The range of this metric is a number from 0 to 100. a higher score indicates easier reading. An average document has an FKI score between 6 - 70. As a rule of thumb, scores of 90-100 can be understood by an average 5th grader. 8th and 9th grade students can understand documents with a score of 60-70; and college graduates can understand documents with a score of 0-30.

D. *The Gunning's Fog Index (GFI):*
Another readability metric used in this study is the Gunning's Fog Index, which was originally proposed by Robert Gunning [23]. It uses average length of sentences and hard word's percentage. The equation for calculating Gunning's Fog Index is as follows:

*GFI = 0.4 (ASL + PHW)*

where, ASL is Average Sentence Length and PHW is Percentage of Hard Word. This metrics is similar to the Flesch scale in that it compares syllables and sentence lengths. A GFI score of 5 means readable, 10 means hard, 15 means difficult, and 20 indicates very difficult. Based on its name, 'Foggy' words are words that contain 3 or more syllables.

E. *Coleman-Liau Index (CLI):*
Meri Coleman and T. L. Liau [25] defined another readability index like ARI, which is used in this study to determine readability value of the code. It is calculated using following formula:

*CLI = 0.0588L – 0.296S – 15.8*

where, L and S are Average number of Letters and Sentences respectively. It relies on characters instead of syllables per word and sentence length. A CLI value computed using the aforementioned formula signifies a high school grade. For example, CLI = 10.6 for a given text means, the text is appropriately readable for a 10-11th grade high school student.

F. *Buse Readability Score (BRS):*
Buse Readability Score is a probability for a given piece of code being highly readable. Buse and Weimer [14] developed a two-class (i.e., high readable, low readable) machine learning classifier based on a number of program constructs as features. BRS is the probability that a given piece of code is classified into the high readable class.

## 3.2 Complexity Metrics

In this study, we have chosen Halstead Complexity Volume and McCabe's Cyclomatic Complexity to as software complexity metrics.

A. *Halstead Complexity Volume:*
Halstead complexity or Halstead volume was introduced by Halstead in 1977 [18]. It is calculated using the following formula:

$$Halstead\ Volume = N * log_2 n$$

having N = $N_1 + N_2$, and n = $n_1 + n_2$ where, $N_1$ and $N_2$ are total number of operators and operands respectively while

$n_1$ and $n_2$ are the number of *distinct* operators and *distinct* operands respectively.

B. *McCabe's Cyclomatic Complexity:*
McCabe's cyclomatic complexity [19], denoted as *M*, is calculated using Control Flow Graph (CFG) where the complexity of data flow remains ignored. It is calculated using following formula.

$$M = E - N + 2P$$

where, *E* = the number of edges in the CFG,
N = the number of nodes in the CFG,
P = the number of connected components in the CFG.

## 3.3 Tools and Datasets

For computing the aforementioned *readability* metrics, we use two tools. The first tool we use is Buse & Weimer readability tool [14] to compute the readability score of our dataset. It is a command line java executable. The second tool [29] is an open-source web application, basically a parser, to determine ARI, SMOG, Gunning's Fog index, Flesch-Kincaid readability index, and Coleman-Liau index.

For computing the complexity metrics (i.e., Halstead Volume and McCabe's cyclomatic complexity), we use JHawk, GUI based application.

To determine the existence of statistically significant correlation between metrics, we use the Pearson Product Moment Correlation Coefficient [30]. In addressing the second research question (RQ2), we use WEKA [26], which is tool that comes with implementation of a collection of machine learning algorithms. All the tools, for this work, are selected based on their accuracy, popularity, and availability.

In our study, we use the same dataset as of the work of *Tashtoush* et a. [7]. The dataset contains 35 different Java programs covering 23 distinct code constructs (e.g., loop, nested loop, conditionals).

## 4. Analysis and Findings

We carry out our analysis in two stages organized in accordance with our two research questions outlined in Section 1.

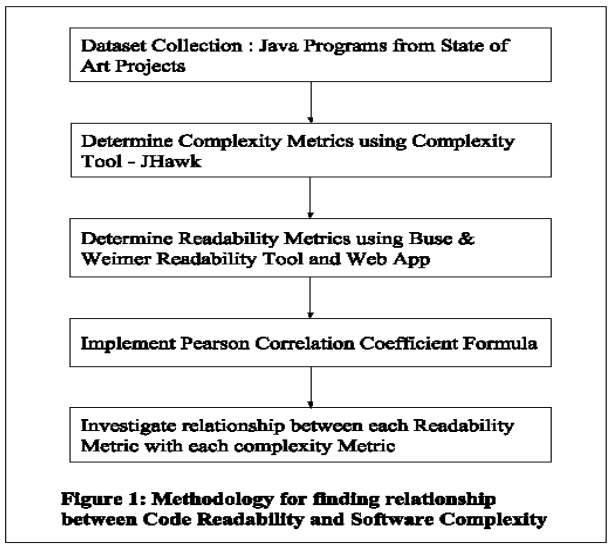

Figure 1: Methodology for finding relationship between Code Readability and Software Complexity

### 4.1 Readability and Complexity

In Figure 1, we summarize the steps involved in addressing our first research question RQ1, which includes the investigation of the relationship between code readability and software complexity. Using the two readability tools mentioned before, we compute all the readability metrics (as introduced in Section 3.1) for each of the 35 Java programs in our dataset. Similarly, using JHawk, we compute the two-complexity metrics (introduced in Section 3.2) for each of the 35 Java programs. Then we investigate if there exists any correlation between the readability and complexity metrics.

To examine the possible correlation, we use Pearson product-moment correlation coefficient [30], which is a well-established statistical measurement to examine linear relationship between variables. We chose to use Pearson coefficient because it is suitable for interval data. The Pearson product-moment correlation coefficient $r_{xy}$ between variables x and y is calculated by:

$$r_{xy} = \frac{\sum_{i=1}^{n}(x_i - \bar{x})(y_i - \bar{y})}{\sqrt{\sum_{i=1}^{n}(x_i - \bar{x})^2 \sum_{i=1}^{n}(y_i - \bar{y})^2}}$$

where $x_i$ and $y_i$ are values of the variables x and y, n is the number of samples (values) available for those variables, $\bar{x}$ and $\bar{y}$ are respectively the mean of all n values of x and y.

The value of $r_{xy}$ ranges between +1.0 and −1.0 and indicates to what extent the variables are positively or negatively correlated. Two variables are positively correlated if one increases, then the other also increases. Negative correlation between variables implies if one gets larger, then the other gets smaller. Positive value of $r_{xy}$ implies positive correlation and negative value implies negative correlation. The closer $r_{xy}$ to ±1.0 the stronger the correlation relationship is. A value of $r_{xy}$ close to zero indicates very weak or no correlation between the variables.

Table 1: Pearson correlation coefficients between readability metrics and complexity metrics

| Readability Metrics (x) | Pearson Correlation Coefficients ($r_{xy}$) | |
|---|---|---|
| | Halstead Volume Complexity (y) | McCabe's Cyclomatic Complexity (y) |
| ARI | -0.172 | -0.115 |
| BRS | -0.315 | -0.248 |
| CLI | -0.503 | -0.425 |
| GFI | -0.210 | -0.136 |
| FKI | -0.163 | -0.172 |
| SMOG | -0.411 | -0.308 |

Table 1 presents the correlation coefficients between each readability metric and each complexity metric. For example, there is a negative correlation with correlation coefficient -0.136 between cyclomatic complexity and the GFI (Gunning's Fog Index) readability metric. As seen in Table 1, each of the readability metrics is negatively correlated with both the complexity metrics. We therefore, derive the answer to our first research question RQ1 as follows.

> *Ans. to RQ1:* There exists a negative correlation between code readability and program complexity. This means that low readability increases program complexity while high complexity degrades code readability.

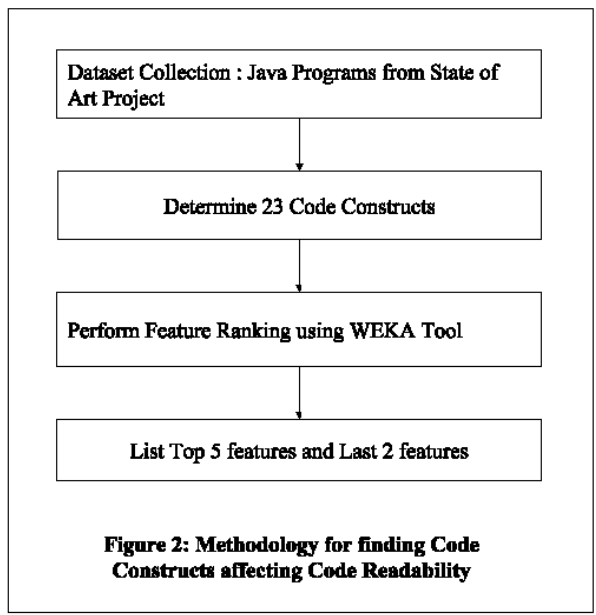

Figure 2: Methodology for finding Code Constructs affecting Code Readability

## 4.2 Code Constructs Affecting Readability

In Figure 2, we summarize the procedural steps involved in addressing our second research question RQ2, which include an analysis for identifying those code constructs that substantially affect code readability. For the purpose, we use dataset provided by the authors of "Impact of Programming Features on Code Readability" [7]. The dataset contains 23 different code constructs and their values for 35 Java programs. A list of the 23 code constructs is presented in Table 2.

Table 2: The 23 distinct code constructs in the dataset

| 1. Lines of Code (LOC) | 13. Line Length distribution |
|---|---|
| 2. Meaningful Names | 14. Identifier name Length |
| 3. Comment Indents | 15. Identifier frequency |
| 4. Indents | 16. IF-else |
| 5. Scope | 17. Nested if |
| 6. Inheritance | 18. For Loop |
| 7. Polymorphism | 19. While Loop |
| 8. Class Distribution | 20. Do-While Loop |
| 9. Spacing | 21. Nested -loop |
| 10. Recursive | 22. Switch |
| 11. Formulas | 23. Array |
| 12. Consistency | |

To identify the code constructs that affect readability, we apply machine learning using the WEKA [26] tool. In the machine learning context, each Java program is considered as an instance and each code construct is regarded as a feature.

The dataset has readability values determined by the authors in the paper [7] which is denoted as $R_P$, for each Java program $P$ in the dataset. Then we manually classify and label each program $P$ into high or low readability depending on its readability score $R_P$ using the following formula:

$$readability_P = \begin{cases} high, & if\ R_P \geq k \\ low, & if\ R_P < k \end{cases}$$

Here, $k$ is the threshold that defines the boundary between the classes. In our work, $k = 12$. Upon completion of labeling the programs in accordance with their high or low readability, we then invoke WEKA to perform feature extraction and ranking.

For feature ranking with WEKA, we choose the correlation-based feature selection, because it is a popular technique for selecting the most relevant attributes/features in a given dataset. WEKA supports correlation-based feature selection by the "CorrelationAttributeEval" technique that requires Ranker search method. We thus determined the strength and direction of correlation between each of the code constructs with code readability. The features (i.e., code constructs) and their rankings obtained this way are presented in Table 3.

The code constructs in Table 3 are ranked and ordered in accordance with their strength of positive correlation with code readability. For brevity, in Table 3, we include the top (strongest) five and bottom (weakest) two code constructs that are positively correlated with readability.

The results in Table 3 are meant to be interpreted as follows. The presence of comments, proper spacing, meaningful names contribute in high readability of code. While loop is better than do-while loop for better readability. Uses of arrays and nested loop do not much help in program readability.

Table 3: Code constructs that affect code readability

| Code Constructs | Rank | Average Merit |
|---|---|---|
| Comments | 1 | 0.53 +- 0.048 |
| Spacing | 2 | 0.496 +- 0.031 |
| While Loop | 3 | 0.361 +- 0.026 |
| Meaningful Names | 4 | 0.312 +- 0.048 |
| Do while loop | 5 | 0.248 +- 0.032 |
| … | … | … |
| Array | 22 | 0.048 +- 0.04 |
| Nested-loop | 23 | 0.047 +- 0.042 |

Based on the findings in Table 3 and the discussion above, we now derive the answer to our second research question RQ2 as follows:

> *Ans. to RQ2:* The five strongest code constructs that positively affect code readability are the presence of comments, spacing, while loop, meaningful names and do-while loop. The use of nested loop and arrays do not much help in increasing readability.

## 5. Threats to Validity

The findings from this study are derived from an analysis of only 35 small programs written in Java. Thus, the results may not be generalizable to programs written in other programming languages and large industrial systems.

Although six readability metrics are used to assess readability of code, they may not be conclusive, since readability is something subject to human perception and dependent on human expertise.

The methodology of this study including the procedure for data collection and analysis is documented in this paper. The dataset is available online [31] and the tools used in this study are also publicly available. Therefore, it should be possible to replicate this study.

## 6. Conclusion

Code Readability and Software Complexity have significant impact on the software quality. Many researchers have investigated on the measurement of code readability and invented tools and metrics to have a standard judgement on the code readability and software complexity. In this paper, using six code readability metrics and two software complexity metrics, we empirically derived evidence of inverse correlation between code readability and complexity. Applying a machine learning technique, we also identified those code constructs that substantially affect readability. The findings from the study are derived from analyses of 35 java programs covering 23 distinct code constructs. The top five code constructs that positively affect readability are comments, spacing, while loop, meaningful names and do-while loop.

In future, we plan to extend this work including programs of different sizes and written in diverse programming languages to gain better viewpoint on the relationship between code readability and software complexity.